\documentclass{ws-procs9x6}

\def\beq{\begin{equation}}
\def\eeq{\end{equation}}
\def\bea{\begin{eqnarray}}
\def\eea{\end{eqnarray}}

\def\gappeq{\mathrel{\rlap {\raise.5ex\hbox{$>$}}
{\lower.5ex\hbox{$\sim$}}}}
\def\lappeq{\mathrel{\rlap{\raise.5ex\hbox{$<$}}
{\lower.5ex\hbox{$\sim$}}}}

\def\snu{\tilde{\nu}}
\def\Huv{\langle H_u \rangle}

\begin{document}

\title{ Parametrizations of the Seesaw \\ 
\lowercase{or} \\
Can the Seesaw be tested?} 

\author{ Sacha Davidson
\footnote{work 
supported by a \uppercase{PPARC} \uppercase{ A}dvanced 
\uppercase{F}ellowship}}

\address{Dept of Physics, University of Durham,
Durham, DH1 3LE, England\\
E-mail: sacha.davidson@durham.ac.uk}

\maketitle

\abstracts{
This proceedings 
contains a review, followed by a more speculative discussion.
 I review different
coordinate choices on the 21-dimensional  parameter
space of the 
 seesaw,
and which of these 21 quantities are observable. 
In MSUGRA, 
there is a 1-1 correspondance between the parameters, and 
the interactions of light (s)particles. 
However,  not all of the 21  can
be extracted
from data,
so the answer to the title question  is ``no''.
How to parametrise the remaining unknowns
is confusing---different choices seem to give
contradictory results (for instance, to
the question ``does the Baryon Asymmetry depend
on the CHOOZ angle?'').
I speculate
on possible resolutions of the puzzle.
}

\section{Introduction}

The seesaw mechanism \cite{seesaw} is a theoretically elegant way
to get the small neutrino masses we observe. It predicts that
the light neutrino masses are majorana, which could be
verified in neutrinoless double $\beta$ decay experiments.
In the absence of Supersymmetry, it predicts that
lepton flavour violation (LFV), and CP violation are suppressed
by powers of the neutrino mass, making the rates very low
outside the neutrino sector.
On the other hand, if spartners were discovered, for instance
at the LHC, observable CP and flavour violation can
be imprinted by the seesaw into the slepton mass matrices.
Experimentally verifying these predictions would
increase our confidence in the seesaw.
Measuring something different---for instance 
majorana masses, no SUSY, and large  neutrino
magnetic moments---would indicate that there is 
other new physics, or  more new physics
in the lepton sector than just the seesaw (see {\it e.g.}
Smirnov, in this volume).
The aim here is to ask if
 we can  {\it test} the seesaw, or as discussed below,
a particular implementation of the seesaw mechanism.

This proceedings is written from a bottom-up phenomenological perspective.
I want to make as few assumptions as possible about
the theory at scales above $m_W$, so I assume the particle
content is the Standard Model (SM), or the  MSSM with universal
soft masses, plus three $\nu_R$,
and allow all possible renormalisable  interactions.
This gives the Lagrangian (in the SM case)
\beq
{\it L} = { Y_e} \bar{e}_R H_d \cdot \ell_L + 
{ Y_\nu}  \bar{\nu}_R H_u \cdot \ell_L + \frac{ M}{2} \bar{\nu^c}_R \nu_R + h.c.
\label{1}
\eeq
where $\ell$ are the lepton doublets, 
$\bar{\nu^c}_R = (C \nu_R^*)^\dagger \gamma_0$, and generation
indices are suppressed. The
 index order on the Yukawa matrices  is right-left.

To test this implementation of the seesaw mechanism, we need to
\begin{enumerate}
\item extract the unknown parameters of eqn (\ref{1})  from data
\item predict an additional observable calculated from those parameters
\item verify the prediction
\end{enumerate}
These proceedings discuss the first step. If it could
be accomplished successfully,  we could
 calculate the baryon asymmetry produced
in various leptogenesis\cite{FY,thermal,others} mechanisms 
(see  Hambye and Raidal  in this volume), which   
would be a fabulous cross-check of
particle physics and cosmology.

There are many other versions of the seesaw (2 $\nu_R$,
type II with scalar triplets, with extra singlets...), which are
motivated  from various theoretical
perspectives (see T Hambye in this volume). 
The model used here contains three $\nu_R$ because
there are three generations, and only three $\nu_R$ because
it is useful to know how well the simple model works before
 adding complications.

I want to test the seesaw {\it mechanism}, rather than
a particular model, so
GUT models,  textures, and theoretical considerations
of ``naturalness''  are avoided (insofar as possible).
The seesaw  mechanism { can} accomodate
{\it any} neutrino masses and mixing angles
(And almost any sneutrino mass matrix\cite{di1}).
Particular  models may prefer certain
ranges for observables, so  data  can provide hints about the theory
that gives the Lagrangian of eqn (\ref{1}). This is
discussed elsewhere in this volume (G Ross and
P Ramond). However, if these theoretical  expectations are not fulfilled,
it is difficult to know if the  model
was wrong, or if there is more new
physics in addition to the seesaw mechanism.

\section{Parametrisations}

Twenty-one parameters are required\cite{21} to fully determine the
Lagrangian of eqn (\ref{1}). Three
of the possible ways these can be chosen are 
discussed here.

 The  usual {\bf ``top-down''}
description of the theory is as follows.
At energy scales $\Lambda \gappeq M$,
where the $\nu_R$ are propagating degrees of freedom, 
one can always choose  the 
$\nu_R$  basis where the mass matrix $M$ is diagonal,
with positive and real eigenvalues:
$M = D_M$.
Similarly, one can choose the $\ell$  basis such
that the charged lepton Yukawa $Y_e$ is diagonal on
its LH indices:  $Y_e^\dagger Y_e = D_{Y_e}^2$.
The remaining neutrino Yukawa matrix $Y_\nu$ is an arbitrary
complex matrix, from  which three
phases can be removed by phase redefinitions on
the $\ell_i$.  It is therefore  described by 
9  moduli and 6 phases, giving
in total 21 real parameters  for the seesaw. 
See \cite{21} for a more elegant counting,
in particular of the phases.

To relate various parametrisations of the seesaw,
it is useful to diagonalise $Y_\nu$, which
can be done with  independent
unitary transformations on the left and right:
\beq 
Y_\nu = V_R^\dagger D_{Y_\nu} V_L
\eeq
So in the top-down approach, the lepton sector can be  described by
the nine  eigenvalues of $D_M , D_{Y_\nu}$ and $D_{Y_e}$,
and the six angles and six phases of $V_L$ and
$V_R$. 
 Notice that in this parametrisation,
 the inputs  are masses and coupling
constants of the  propagating particles at energies
$\Lambda$,  so it makes ``physical'' sense.

The effective mass matrix $m$  of the light neutrinos can be
calculated, in the $D_{Y_e}$ basis (charged lepton mass 
eigenstate basis):
\beq
m \equiv \kappa \Huv^2  =   Y_\nu^T D_M^{-1} Y_\nu \Huv^2
  =  V_L^T  D_{Y_\nu} V_R^* D_M^{-1} V_R^\dagger D_{Y_\nu} V_L  \Huv^2
\eeq
$\kappa$ is introduced to avoid  the  Higgs vev $\Huv$
cluttering up formulae. The leptonic mixing matrix $U$
is extracted by diagonalising $\kappa$:
\beq
\kappa = U^* D_\kappa U^\dagger
\eeq
where $D_\kappa = diag  \{\kappa_1, \kappa_2, \kappa_3 \}$, and
$U$ is parametrised as
\bea U= \hat{U}\cdot {\rm diag}(1 ,e^{i\alpha} ,e^{i\beta}) ~~~.
\label{UV}
\eea
 $\alpha$ and $\beta$ are ``Majorana'' phases,
 and $\hat{U}$ has the form of the CKM matrix
\beq
 \label{Vdef} 
\hat{U}= \left[ \begin{array}{ccc}
c_{13}c_{12} & c_{13}s_{12} & s_{13}e^{-i\delta} \\
-c_{23}s_{12}-s_{23}s_{13}c_{12}e^{i\delta} & 
 c_{23}c_{12}-s_{23}s_{13}s_{12}e^{i\delta} & s_{23}c_{13} \\
s_{23}s_{12}-c_{23}s_{13}c_{12}e^{i\delta} & 
 -s_{23}c_{12}-c_{23}s_{13}s_{12}e^{i\delta} &
  c_{23}c_{13}  
\end{array} \right]  ~~~.
\eeq

Alternatively, the (type I) seesaw Lagrangian of eqn (\ref{1}) 
can be described with inputs from the left-handed sector \cite{di1}.
This is refered to as a {\bf ``bottom-up''} parametrisation, because
the left-handed (SU(2) doublet) particles have masses
$\lappeq$ the weak scale. 
$D_{Y_e}$,  $U$ and  $D_{\kappa}$,
can be taken as a subset of the inputs.
To identify the remainder,
imagine sitting in  the $\ell$ basis 
where $\kappa$ is diagonal, so as
to emphasize the parallel between this
parametrisation and the previous one
(this is similar to the $\nu_R$ basis
being chosen to diagonalise $M$). If one knows
$Y_\nu^\dagger Y_\nu \equiv W_L D_{Y_\nu}^2 W_L^\dagger$ 
in the $D_{\kappa}$ basis,  then the $\nu_R$
masses and mixing angles can be calculated:
\beq
 M^{-1}   = 
 D_Y^{-1} W_L^* D_\kappa W_L^\dagger D_Y^{-1}  
  =   V_R^* D_M^{-1} V_R^\dagger  
\eeq

In this parametrisation, there are three possible
basis choices for the  $\ell$ vector space: the
charged lepton mass eigenstate basis ($D_{Y_e}$),
the neutrino mass eigenstate basis ($D_\kappa$), 
and the basis where the $Y_\nu$ is diagonal. The first
two choices are physical, that is, $U$ rotates between
these two bases.
$D_{Y_e}$,$D_\kappa$ and $U$  contain the 12 possibly measurable parameters
of the SM seesaw. The remaining 9 parameters can be
taken to be $D_{Y_\nu}$  and $V_L$ (or $W  = V_L U$).
In SUSY one can hope to extract these parameters from
the slepton mass matrix.

The {\bf Casas-Ibarra} \cite{CI} parametrisation is very
convenient for calculations. It uses
 ${ D_M}$,  ${ D_\kappa}$ and $D_{Y_e}$ as inputs,
and  the transformations $U$ and $R$, which
go  between the bases where these matrices are diagonal. 
$U$ is the usual leptonic mixing matrix.
The matrix $R = D_M^{-1/2} Y_\nu D_{\kappa}^{-1/2} $, 
 is a complex {\it orthogonal} matrix, which
transforms between
the  ${ D_M}$ and  ${ D_\kappa}$ bases. (Since
$M$ and $\kappa$ are respectively in the RH and LH
neutrino vector spaces, it is unsurprising that the
transformation matrix is not unitary.)
R can be written  as 
$R = {\rm diag} \{ \pm 1, \pm 1, \pm 1 \} \hat{R}$
where  the $\pm 1$ are related to the CP parities of the $N_i$,
and  $\hat{R}$ is 
an orthogonal matrix with complex angles:
\beq
\hat{R} =  \left[ \begin{array}{ccc}
c_{13}c_{12} & c_{13}s_{12} & s_{13} \\
-c_{23}s_{12}-s_{23}s_{13}c_{12} & 
 c_{23}c_{12}-s_{23}s_{13}s_{12} & s_{23}c_{13} \\
s_{23}s_{12}-c_{23}s_{13}c_{12} & 
 -s_{23}c_{12}-c_{23}s_{13}s_{12} &
  c_{23}c_{13}  
\end{array} \right]  ~~~.
\label{R}
\eeq

The aim of this proceedings is to reconstruct the RH seesaw parameters from the
LH ones, many of which are accessible at low energy. 
However,as discussed in the following section, 
reconstruction is impossible.
We can at best  try to establish relations between observables, which
turns out to be quite confusing. 
$R$ will be helpful in  discussing these puzzles.

In summary, the lepton sector of the SM + seesaw can be parametrised with
$D_{Y_e}$,  the real eigenvalues of
two more matrices, and the transformations
among the bases where the matrices
are diagonal.  The  matrices-to-be-diagonalised can
be chosen in various ways:
\begin{enumerate}
\item `` top-down''---input the $\nu_R$ sector: ${ D_M}$, 
${ D_{Y_\nu Y_\nu^\dagger}},$ and ${ V_R}$ and $V_L$.
\item `` bottom-up''---input the $\nu_L$ sector:  ${ D_\kappa}$, 
${ D_{Y_\nu^\dagger Y_\nu}},$ and ${ V_L}$ and $U$.
\item ``intermediate''---the Casas-Ibarra parametrization: 
${ D_M}$, ${ D_\kappa}$, and $U$ and a complex orthogonal matrix ${ R}
$. 
\end{enumerate}

\section{(Supersymmetric) reconstruction?}

If  the matrices
 ${ D_\kappa}$, $D_{Y_e}$,
${ D_{Y_\nu^\dagger Y_\nu}},$  ${V_L}$ and $U$ were
known, it would be possible to
reconstruct the masses and mixing angles of the 
$\nu_R$. Can  the elements of
these matrices be determined \cite{di1}?

We know the masses of the charged leptons, so we
know $D_{Y_e}$ (modulo $\tan \beta$ in SUSY models).

We know two mass differences in the neutrino sector.
If the light neutrinos are degenerate, measuring 
the overall scale of  their masses is possible and
would determine $D_{\kappa}$. However, if
the mass pattern is hierarchical or inverse hierarchical,
we would know only $\kappa_3$ and $\kappa_2$.
See the contribution of K Heeger, for  present and
future accuracy on $D_\kappa$, and $U$.

In the mixing matrix $U$, we currently know
two angles. We hope to measure the third, and
also the ``Dirac'' phase $\delta$. But the
``majorana'' phases appear only in slow lepton
number changing processes, so at the moment do
not seem experimentally accessible \cite{nogo}. 

The remaining parameters to be determined are
the eigenvalues of $Y_\nu$, and the  matrix $V_L$.
In supersymmetric models 
$Y_\nu$ 
contributes via loops to the slepton mass matrix.
Consider a model, such as gravity-\cite{borz} or anomaly-mediated
\cite{a-m}
SUSY breaking, where the soft masses are universal
at a scale $\Lambda >M_3$. In renormalisation group
running between $\Lambda$ and $m_W$, the slepton
mass matrix will acquire flavour off-diagonal terms,
due to  loops involving the $\nu_R$ 
(see  Masiero in this volume). Using the leading log approximation
for the RG running, the sneutrino mass  matrix, in
the $D_{Y_e}$ basis, is:
\bea
\label{softafterRG} 
\left[ m^2_{ \snu}\right]_{ij} & \simeq & 
 \left({\rm diag\,\, part}\right)
-\frac{3m_0^2 + A_0^2}{8\pi^2}
({\bf Y^{\dagger}_\nu})_{ik} ({\bf Y_\nu})_{kj} \log\frac{\Lambda}{M_k} 
\eea
where $m_0$ and $A_0$ are the universal soft parameters
at scale $\Lambda$.  

It is tantalising that the seesaw contribution
to flavour violation in the sleptons is potentially
observable, and depends on
the heavy neutrino masses in a different way than
$\kappa$.  
{\it If} we could determine $\left[ m^2_{ \snu}\right]$
exactly (the three masses, three mixing angles, and
three phases), and 
{\it if} we take seriously the assumption of universal
soft masses, 
then we could reconstruct the renormalisable interactions
of the high-scale seesaw---that is, the $\nu_R$ masses and
Yukawa couplings---from the mass matrices and mixing
angles of weak-scale particles. 

Unfortunately, neither of these conditions is likely to be fulfilled.
Firstly, not all the parameters of  $\left[ m^2_{ \snu}\right]$
can be measured with the required accuracy.
The diagonal elements of the second term of
eqn (\ref{softafterRG}) shift $ m^2_{ \snu}$ by 
of order $y_i^2  \%$, so
a large $Y_\nu$ eigenvalue $\sim 1$ could have
a measurable effect. However, if $Y_\nu$ has a hierarchy
similar to the quark Yukawas, the effects of the first
and second generation $y_i$ are (undetectably)  small.

The flavour-changing elements of eqn (\ref{softafterRG}) 
could be seen   
at colliders \cite{feng2}, and 
induce   rare decays, such as $ \mu \rightarrow  e \gamma$
\cite{meg}.
A very optimistic  experimental sensitivity of order
$BR(\tau \rightarrow \ell \gamma)  \sim 10^{-9}$  
(the current limit is $\sim 10^{-7}$), 
could probe  $|[V_L]_{3 \ell}[V_L]_{3 \tau}^* y_3^2| \gappeq 10^{-(1 \div 2)}$.
$\mu-e$ flavour violation is  more encouraging:
there are plans to reach  $BR( \mu \rightarrow  e \gamma) \sim 10^{-13}$,
which would be sensitive to 
$|[V_L]_{3 e}
[V_L]_{3 \mu}^* y_3^2| \gappeq 10^{-(3 \div 4)}$.
However, to extract a ``measurement'' 
of  either of the $|[V_L]_{3 \ell}|$ from 
rare decays would require knowing all the 
masses and mixing angles for the other SUSY
particles contributing to the decay.

For hierarchical $Y_\nu$ eigenvalues,
eqn (\ref{softafterRG}) implies that the
three off-diagonal elements of $\left[ m^2_{ \snu}\right]$,
are determined by two matrix elements of $V_L$.
So one angle of $V_L$ is unknown, 
and there should be some correlation between
 $\left[ m^2_{ \snu}\right]_{\tau \mu}$,
$\left[ m^2_{ \snu}\right]_{\tau e}$,and
$\left[ m^2_{ \snu}\right]_{\mu e}$.
Notice, however, that this is a prediction
of hierarchical $Y_\nu$. 
In the bottom-up parametrisation,
the slepton mass matrix  determines $V_L$ and $D_{Y_\nu}$,
rather than the seesaw making predictions for
 $\left[ m^2_{ \snu}\right]$.

Now we come to the three phases of $V_L$. To
extract all of these is quite hypothetical; it 
 would require three independent
measurements of CP violation in the sleptons.
Two possibilities at colliders are
charged lepton asymmetries in  slepton decays
\cite{feng}, and sneutino-anti-sneutrino oscillations
\cite{haber}. The slepton phases also contribute
to CP violating observables in the leptons, in
conjunction with phases from other SUSY particles. 
This is discussed in this volume by Hisano.

The second objection
to extracting  seesaw parameters from
eqn (\ref{softafterRG}), is that
we do not know that  soft masses are universal.
It is a reasonable assumption in top-down analyses, because
we know that flavour violation mediated by sparticles
must be suppressed.  But I know of no way to distinguish 
contributions to  $\left[ m^2_{ \snu}\right]$  that come from
the RG running with the seesaw, from those that come from  non-universal
soft masses, threshhold effects, other particles with flavour
off-diagonal couplings, etc... So measuring
$\left[ m^2_{ \snu}\right]$ exactly could be used to set
an upper bound on the seesaw contributions (if one
makes  the reasonable
assumption that there are no cancellations among different
contributions), but would not determine them.

It is also possible-in-principle
to reconstruct the {\bf non-SUSY } seesaw: 
Broncano $ et~ al.$ \cite{broncano} observed that the
21 parameters can be extracted from the coefficients of dimension
5 and 6 operators in the Standard Model. However,
the coefficients of the dimension 6, lepton number conserving
operators  are suppressed by two powers of the $\nu_R$ mass,
so are (unobservably) small.


In summary, the parameters of the type I seesaw
cannot be extracted from data. 
This should hardly be surprising---we do not usually
expect to reconstruct high-scale theories  ($e.g.$
 which GUT, and how does it break?) from
weak-scale observations.  So why do we even ask if it is possible
in the seesaw? 
I am aware of  two peculiarities,
 which make the seesaw ``reconstructable in principle'':
 the $\nu_R$ 
only have  interactions with light particles (via $Y_\nu$),
and  the effective operators induced at low
energy  are experimentally accessible (in principle!) 
for all flavour indices.
To see why these features are
significant, compare to proton decay---
which I assume to be mediated by a ``triplet
higgsino'' dressed with a squark loop. However
accurately we mesure every available
proton decay channel, we cannot determine
the mass and couplings of the triplet higgsino,
because we must always sum over squark flavours in the loop,
and we only measure proton decay with first
generation quarks in the initial state (unlike
the three generation $\nu$ and $\snu$ mass matrices).

\section{Independence, orthogonality and relations
when we cannot reconstruct}

The 21 parameters of the seesaw cannot be determined
from observation, but 
some sort of partial reconstruction, 
using the available data, could
be possible. This turns out to be much more
confusing than one would anticipate.
To identify the problem, 
imagine calculating
the baryon asymmetry  as a function of parameters
separated into three categories: those
we know now, those we hope to know, and those we will never know. 
It  then seems  straightforward to 
study how the asymmetry depends on, for
instance, $\theta_{13}$. 
But in practise it is anything but transparent (see eqn \ref{eps-dep}):
the  asymmetry is  independent of 
$U$
in the Casas-Ibarra parametrisation, but 
does depends on $U$ in the bottom-up  version.
That is, the choice of parametrisation
for the unmeasurables, changes the dependence of 
one observable (the baryon asymmetry) on another ($\theta_{13}$).
It would be better to ask ``is $\epsilon_1$ {\it sensitive} to
$\theta_{13}$?''---this has a unique and useful answer, as discussed
in the next section.

 The aim of this section is to explore
how different coordinates on seesaw parameter space  depend
on each other, and 
what we mean by ``depends on''  and ``independent'' .
I start by reviewing some contradictory statements
which can be derived using various parametrisations.
Then I present a toy model
using parametrisations of the plane, where
these same contradictions arise, and where
the resolution is obvious. Lastly I suggest how
the analogy of the plane could be related to the
seesaw.

It has been claimed in various papers that  $\epsilon_1$,
the CP asymmetry of thermal leptogenesis, is independent
of the leptonic mixing matrix $U$.  This seems intuitively
reasonable, because leptogenesis involves the 
$\nu_R$, and is independent of $Y_e$. 
 In the limit of
 hierarchical $\nu_R$:
\bea
\epsilon_1 & \simeq& -\frac{3M_1}{8 \pi [Y_\nu Y_\nu^\dagger]_{11}}
\Im \{ Y_\nu \kappa^* Y^T \} \label{10} 
= 
-\frac{3M_1}{8 \pi} \frac{
\Im \{ R^2_{1j} \kappa_j^2 \}}{ |R_{1k}|^2 \kappa_k} \\
& \propto &  \Im \left\{  V_L U D_\kappa^3 U^T V_L^T  D_{Y_\nu}^{-2}
V_L^* U^* D_\kappa  U^\dagger V_L^\dagger D_{Y_\nu}^{-2}    \right\}
\label{eps-dep}
\eea
where the second equality of (\ref{10}) is in
the Casas-Ibarra parametrisation. 
To translate eqn (\ref{10}) into  bottom-up
coordinates,  requires calculating the mass
and eigenvector (first colomn of $V_R$) of $\nu_{R1}$,
which gives short analytic formulae in some limits.
 However,
for hierarchical $\nu_R$ (the limit in which eqn (\ref{10}) is valid),
$\epsilon_1$ is proportional to a Jarlskog invariant\cite{ryu},
which gives eqn (\ref{eps-dep}). 
We see that  $U$ does not appear in the  expression for $\epsilon_1$
in Casas-Ibarra, but does appear 
in the bottom-up parametrisation. 
So it is unclear whether  $\epsilon_1$ depends on $U$---what do we 
mean by ``depend''?
If  a mathematical definition
 can  be constructed, then there should be a unique answer.

We can draw an analogy between coordinate choices on a manifold,
 and parametrisation choices for the seesaw. 
Different coordinate choices on the plane,
all  used with the same
\footnote{Of course,  we know that this is wrong; the metric should
change with the coordinate system.}
  metric $\delta_{\alpha \beta}$, give 
confusing results that ressemble  the puzzle 
about whether $\epsilon$ depends on $\theta_{13}$.
If we use the appropriate  metrics, results are independent
of the coordinate system,
which  can be chosen for calculational
convenience.  There is no metric given on ``seesaw parameter space'',
 but this analogy suggests that inventing one
would resolve the  confusion.

Consider two 
choices of  coordinates on the upper half plane:
\begin{enumerate}
\item the cartesian $(y,z)$ with $ y > 0$ and 
 metric $g_{\alpha \beta} =  I$.
\item  $ R = \sqrt{y^2 + z^2}$ and $ Z = z$,
with  $R > Z$ and  metric  
\beq
g_{A B} = \frac{R^2}{R^2 - Z^2}\left[ \begin{array}{cc} 1 & -Z/R \\ 
-Z/R &1 
\end{array}
 \right]
\eeq
\end{enumerate}
These are  equally good coordinate
choices for the same flat 2-d surface. The seesaw
analogy  we
want to address is: does $R$ ``depend'' on $Z$?

 In any coordinate system,
the coordinates vary independently. So by definition
\beq
\frac{\partial R }{\partial Z}  = 0
\eeq
which could be taken to mean that `` $R$ is independent of $Z$''.
A more intuitive  quantity is the total derivative, or by
analogy with general relativity, 
the change of $ R$, treated as a scalar
function, along the curve of varying $Z$:
\beq
\left( \begin{array}{cc}   
\frac{ \partial R }{ \partial y} &
\frac{ \partial R }{ \partial z}
\end{array}
 \right)
\left[ \begin{array}{cc} 1 & 0 \\ 
0 &1 
\end{array}
 \right]
\left( \begin{array}{c} \frac{ \partial Z }{ \partial y}  \\ 
\frac{ \partial Z }{ \partial z}
\end{array}
 \right) =  \frac{Z}{R}
\label{*}
\eeq
which is the expected answer. Notice that
we need to know how to transform to cartesian coordinates
(equivalently, the metric on $R,Z$ space)  for this calculation.

To summarise, $\epsilon$ and $\theta_{13}$ are functions
of seesaw parameter space, and can be defined such
that 
\beq
\frac{\partial \epsilon ~}{\partial \theta_{13}} = 0
\eeq
by a suitable choice of parameters.
However, a better measure of whether
$\epsilon$ depends on $\theta_{13}$ would be
something like eqn (\ref{*}). To evaluate
this, we need  a metric on seesaw parameter
space {\footnote{ When doing seesaw parameter space scans,
one must choose the distribution of input points in
parameter space. This number density
(``measure on parameter space'') is motivated by some theoretical model
for the origin of seesaw parameters, so is not intrinsic to
the seesaw. Therefore it is   not related to
this  ``metric''.}}.

How to choose this metric?
The top-down parametrisation is the most natural, so
in two generations, the obvious choice is to
take $\{ D_{Y_e},  V_L, D_{Y_\nu}, V_R, D_M \}$ as 
cartesian coordinates. With this metric, it is
straightforward to show that
$\epsilon$ does vary with the angle of the matrix $U$.
(This is simple, because in 2 generations it is easy  to
calculate the angle of $W_L$ in terms of RH
parameters.)
 However, in three generations, ``distance'' on
the unitary transformations should be invariant 
under 
reparametrizations ($e.g. V_R=U_{12} U_{13} U_{23}$ or
$=U_{23} U_{13} U_{12}$),  suggesting a metric
similar to the one for polar coordinates.


It {\it is} clear, from this section,  that the
``dependance'' of one seesaw observable on another
in {\it not} clear. For example, the coordinates on
seesaw parameter space can be chosen such that
either  $\epsilon_1$  is a function of the
MNS matrix, or it is not. This confusion can be 
resolved by inventing a notion of ``orthogonality''
for coordinates, that is, a metric on parameter
space. However, the metric seems an esoteric solution,
and how to find the correct one is not
obvious.

\section{Rethink: what happens in the Standard Model?}

In the Standard Model, the Lagrangian parameters can be
reconstructed from data---in fact,
there are many more measurements than parameters, so 
the SM is tested at part-per-mil accuracy.  But some
parameters are better determined than others, so the
difference with respect to the seesaw is just the size
of the error bars.

In the SM, the key is  the {\it sensitivity} of data to
a parameter. For instance, to determine
$m_t$ from electroweak data, one should choose
an observable   with large $m_t^2$ corrections,
and a parametrisation ($eg$, definition of
$s_W^2$), where these are easy to identify.
If the parameters other than $m_t$ are sufficiently
well determined, a range for $m_t$ can be extracted
\footnote{In reality this is a crude approx to doing a combined fit.}.
This is self-evident;  the data allows 
a model to occupy a subset (often a 
  multi-dimensional ellipse) in parameter space.

We say an observable  $Ob$ is sensitive to a parameter $P$,
if measuring $Ob$ constrains $P$ to sit in a certain range.
Conversely,  $Ob$ is insensitive to $P$, if measuring
$Ob$ is consistent with any value of $P$ (possibly because
one ajusts other unknowns to compensate for variations in $P$).

So returning to the seesaw, 
one could conclude that  ``does $\epsilon_1$ depend
on $\theta_{13}$?'' is the wrong question. If instead,
one asks ``is $\epsilon_1$ {\it sensitive} to $\theta_{13}$?'',
then the answer at present  is clearly no. It is easy to see, 
in the parametrisation using $R$, that any value
of $\theta_{13}$ is consistent with the observed baryon a
symmetry.

\section{Summary}

The seesaw generates small neutrino  masses, 
by introducing   heavy majorana $\nu_R$s, which share
a Yukawa  coupling with the lepton doublets of the 
Standard Model. 
It is theoretically possible to establish a 1-1 correspondance between
observables (in the quantum mechanical sense), and the 21 parameters
of the seesaw (type I, 3 generations).
This correspondance, and two other parametrisations
of the seesaw, are discussed in section 2. 
Unfortunately, this  peculiarity of
the seesaw
does not mean
the parameters can be extracted from data;  some of
the ``observables'' are not realistically measurable, and others
cannot be determined accurately enough (see section 3). This makes
the seesaw mechanism difficult to test, according to
the definition of test outlined in the introduction.

This is sad because the dream test of the seesaw would be to
extract its parameters from data, calculate the baryon asymmetry
produced in leptogenesis---and get the right answer. 
More realistically, we can ask ``is the baryon asymmetry {\it sensitive}
to any of the seesaw's measurable parameters?'' For instance,
does generating the baryon asymmetry by a specific
leptogenesis scenario  imply that 
 $\theta_{13}$, or the phase
$\delta$,  should occupy  restricted ranges?
Again, the answer   sadly seems to be ``no''. 
More generally, one could  study which observables 
are sensitive to which parameters,
 $e.g.$ would $BR(\mu  \rightarrow e \gamma) \neq 0$ restrict the majorana
phases of MNS \footnote{In the Casas-Ibarra
parametrisation, $BR(\mu  \rightarrow  e \gamma)$ 
depends on theses phases
\cite{pet}}?
Most studies to date have looked at whether an  observable $O$ ``depends''
on  a parameter $P$---which is not such
a useful question, because the answer depends on the
parametrisation. Section 4 attempts to construct
a parametrisation-independent definition of ``depend'',
not very successfully. So it is better to
ask if $O$ is $sensitive$ to $P$, which does
have a unique answer, as discussed in  
section 5.

\subsection*{Acknowledgements}
I wish the seesaw many happy returns, and thank
the organisers for a very enjoyable birthday
celebration, with many interesting speakers and participants.
In particular, I thank S Vempati for clarifying questions, 
S Lavignac for discussions and a careful reading of
the manuscript, 
and S Petcov for many
interesting discussions about seesaw parametrisations.

\end{document}